\documentclass[aps,pre, superscriptaddress, twocolumn]{revtex4-1}

\usepackage[dvips]{graphicx}
\usepackage{amsmath}
\usepackage{amssymb}
\usepackage{natbib}
\usepackage{color}
\usepackage{epsfig}
\usepackage{subfigure}
\begin{document}

\title{A unifying model for random matrix theory in arbitrary space dimensions }
\author{Giovanni M. Cicuta}
\email{cicuta@fis.unipr.it}
\affiliation{Dip.  Fisica, Universit\`{a} di Parma,  Parco Area delle Scienze 7A, 43100 Parma, Italy ,}
\author{ Johannes Krausser}  
  \author{Rico Milkus}  
\author{Alessio Zaccone}
\email{az302@cam.ac.uk}
\affiliation{Statistical Physics Group, Department of Chemical Engineering and Biotechnology, 
 and Cavendish Laboratory, University of Cambridge, Cambridge, CB3 0AS, UK}


\begin{abstract}
A sparse random block matrix model suggested by the Hessian matrix used in the study of elastic vibrational modes of amorphous solids is presented and analyzed. By evaluating some moments, benchmarked against numerics, differences in the eigenvalue spectrum of this model in different limits of space dimension $d$, and for arbitrary values of the lattice coordination number $Z$, are shown and discussed. As a function of these two parameters (and their ratio  $Z/d$), the most studied models in random matrix theory (Erdos-Renyi graphs, effective medium, replicas) can be reproduced in the various limits of block dimensionality $d$. Remarkably, the Marchenko-Pastur spectral density (which is recovered by replica calculations for the Laplacian matrix) is reproduced exactly in the limit of infinite size of the blocks, or $d\rightarrow\infty$, which for the first time clarifies the physical meaning of space dimension in these models. The approximate results for $d=3$ provided by our method have many potential applications in the future, from the vibrational spectrum of glasses and elastic networks, to wave-localization, disordered conductors, random resistor networks and random walks.

\end{abstract}


\maketitle

\section{Introduction}
The eigenvalue spectrum of sparse random matrices is a fascinating subject with widespread applications in physics, from the energy levels of nuclei, to random resistor networks, random walks, the electronic density of states of disordered conductors, and many other topics~\cite{handbook}. It was investigated for several decades, from pioneering works \cite{pio} to modern times~\cite{mod}.\\

In particular, random matrix theory has been applied extensively in recent years to the problem of the vibrational spectrum of glasses, where structural disorder leads to a number of puzzling effect in the vibrational density of states (DOS), such as the excess of soft low-energy modes (boson peak) with respect to Debye's $~\omega^{2}$ law~\cite{schirmacher,parisi,bir,vitelli,milkus}. This anomaly in the spectrum is related to  well-know anomalies in the thermal properties at low temperatures~\cite{phillips}.
This remains a famously unsolved problem because its mathematical description is plagued by the impossibility of analytically solving for the eigenvalue spectrum of the Hessian matrix of a disordered solid. 

Recently, replica-symmetry breaking and allied techniques have been applied to the problem of vibrational eigenmodes of glasses, and produced results which recover the well-known Marchenko-Pastur (MP) distribution of eigenvalues of random Laplacian matrices~\cite{parisi}. The big question is about the applicability of these results: both MP and replica are generally thought to be valid for "high-dimensional" systems, but what this means, in practice or in quantitative terms, has remained unanswered. This is clearly a central point of paramount relevance in the current debate on the theoretical description of glasses.

In this work, we clarify for the first time that MP and replica results are exactly valid in the case of random block Laplacian matrix where the dimension of the blocks is infinite. Furthermore, we show that while the lowest eigenvalue of the support is weakly dependent on the space dimension (which ensures that the $\sim(Z-2d)$ scaling of the boson peak frequency in jammed solids and some models of glasses is rather well captured by high-dimensional models~\cite{parisi,beltukov}), instead the shape of the eigenvalue distribution changes significantly with $d$ and therefore high-dimensional methods such as MP and replica may not provide an accurate modelling of the vibrational DOS of disordered solids.

\section{Model}
In all models or random spring networks, the elastic energy is a quadratic function of the displacements of the particles from their instantaneous ``frozen'' positions. The stiffness matrix or Hessian matrix $W$ is a Laplacian random symmetric matrix where each row is comprised of a small and random number of non-zero coefficients. The off-diagonal entries $W_{i,j}$, $i<j$, are identical independent random variables, whereas the diagonal entries $W_{i,i}=-\sum_{j\neq i}W_{i,j}$. The latter requirement is dictated by enforcing mechanical equilibrium on every atom $i$ in the lattice.\\

The most typical model is the study of  the spectrum of the Adjacency matrix or the Laplacian matrix of a Erdos-Renyi graph with $N$ vertices in the limit of large order of the matrices (the large $N$ limit).\\

The only parameter in the model is the probability $p/N$ of a link in the random graph to be present, whereas the dimension $d$ of the space $R^d$ of the amorphous material or the  random spring model is absent.\\

In this work, we consider a block random matrix model which seems the simplest  generalization of the above models, which retains a couple of relevant parameters.\\

We consider a real symmetric matrix $M$ of dimension $N d\times Nd$ where each row or column has $N$ random block entries , each being a $d\times d$ matrix.\\

Every $d \times d$ off-diagonal block has probability $1-Z/N$ of being a null matrix and a probability $Z/N$ of being a rank one matrix, $X_{i,j}=X_{j,i}=(X_{i,j})^t=\hat{n}_{ij}\hat{n}_{ij}^{t}$ where $\hat{n}_{ij}$ is a $d$-dimensional random vector of unit length, chosen with uniform probability on the $d$-dimensional sphere. Furthermore, $\hat{n}_{ij}\hat{n}_{ij}^{t}$ is the usual matrix (or dyadic) product of a column vector times a row vector, which gives a rank-one matrix.

In the formulation of the stiffness matrix $W$, the unit vector  $\hat{n}_{ij}$ provides the direction between vertex $i$ and vertex $j$ (in a disordered solid or elastic network, between two atoms $i$ and $j$). For more details on the Hessian matrix of disordered solids see Refs.\cite{lemaitre,scossa}.\\  
We study two prototypes of such block random matrices called the Adjacency block matrix $A$ and the Laplacian block matrix $L$.\\

\begin{widetext}

\begin{eqnarray}
A=\left( \begin{array}{cccccccc}
0 & X_{1,2} & X_{1,3}& \dots & X_{1,N}\\
X_{2,1}& 0 & X_{2,3} & \dots & X_{2,N}\\
\dots & \dots & \dots & \dots & \dots\\
X_{N,1} & X_{N,2}&X_{N,3}&\dots & 0 \end{array}\right)
  \qquad  \qquad
\label{d.1}
\end{eqnarray}

\begin{eqnarray}
L=\left( \begin{array}{cccccccc}
\sum_{j\neq 1}X_{1,j} & -X_{1,2} & -X_{1,3}& \dots & -X_{1,N}\\
-X_{2,1}& \sum_{j\neq 2}X_{2,j} & -X_{2,3} & \dots & -X_{2,N}\\
\dots & \dots & \dots & \dots & \dots\\
-X_{N,1} & -X_{N,2}& -X_{N,3}&\dots & \sum_{j\neq N}X_{N,j} \end{array}\right)
 \nonumber\\
\label{d.2}
\end{eqnarray}

\end{widetext}

In both the above matrices, the set of $X_{i,j}$, $i<j$ is a set of $N(N-1)/2$ independent identically distributed random matrices and each $X_{i,j}$ is a rank-one matrix and a projector.\\

The study of the spectral density of the matrices $A$, $L$, in the limit $N\to \infty$, with $Z$ fixed and $d$ fixed, is more difficult then the corresponding study with $d=1$, the Erdos-Renyi graph, where all moments of both spectral functions are known \cite{bau}, yet the 
spectral distributions are not known.\\

\section{Evaluation of moments}
Any symmetric matrix $M$ of order $N$ corresponds to a complete graph with $N$ vertices where the non-oriented link $(i,j)$ has the weight $M_{ij}$ and $(M^k)_{ii}$ is evaluated as the sum of the contributions associated to all paths of $k$ steps on the graph from vertex $i$ to itself. We used this familiar technique to evaluate the limiting moments. However in the present case, the contribution of each path is the product of matrices and the evaluation of moments of high order is laborious. We evaluated the first five limiting moments   \\

 \begin{eqnarray}  \mu_k&=&\lim_{N \to \infty}\frac{1}{N\,d} <{\rm Tr}\,A^k>  \quad , \quad \mu_0=1\quad , \quad   \mu_{2k+1}=0 \nonumber\\
	 \nu_k &=&\lim_{N \to \infty}\frac{1}{N\,d} <{\rm Tr}\,L^k> \quad , \quad \nu_0=1 
	\nonumber
	\end{eqnarray}

which produce the following results:

\onecolumngrid
\begin{eqnarray}
\begin{array}{ccccccc}
&\mu_2 = \frac{Z}{d}  \\
&\mu_4 = \frac{Z}{d}&+ 2 \left(\frac{Z}{d}\right)^2 \\
&\mu_6 = \frac{Z}{d} &+ 6\left(\frac{Z}{d}\right)^2 &+5\left(\frac{Z}{d}\right)^3 \\
&\mu_8 =\frac{Z}{d}&+ \left(\frac{Z}{d}\right)^2   \left( 12 +2\, \frac{3}{d+2}\right)  &+28 \, \left(\frac{Z}{d}\right)^3&+14\,\left(\frac{Z}{d}\right)^4  \\
&\mu_{10} = \frac{Z}{d}&+\left( \frac{Z}{d}\right)^2\left( 20+10\frac{3}{d+2}\right)
	& +\left( \frac{Z}{d}\right)^3\left( 90+20\frac{3}{d+2}\right) &+
	 120\, \left( \frac{Z}{d}\right)^4 &+42\,\left( \frac{Z}{d}\right)^5
	\end{array}
\label{d.3}
\end{eqnarray}

\begin{eqnarray}
\begin{array}{cccccccccccc}
&\nu_1 = \frac{Z}{d} \\
&\nu_2 = 2\, \frac{Z}{d} &+ \left( \frac{Z}{d} \right)^2 \\
&\nu_3 = 4\, \frac{Z}{d}& +6\, \left( \frac{Z}{d} \right)^2 &+ \left( \frac{Z}{d} \right)^3  \\
&\nu_4 = 8\, \frac{Z}{d}& + \left( \frac{Z}{d} \right)^2 \left( 24 + \frac{3}{d+2} \right)&+12 \left( \frac{Z}{d}\right)^3 &+\left(\frac{Z}{d}\right)^4  \\
&\nu_5 = 16\, \frac{Z}{d}& + \left( \frac{Z}{d}\right)^2 \left(80  +10\,\frac{3}{d+2}\right)&+\left( \frac{Z}{d}\right)^3 \left(80 +5\,\frac{3}{d+2}\right) &+
20\,\left(\frac{Z}{d}\right)^4 &+ \left(\frac{Z}{d}\right)^5 .
\end{array}
\label{d.4}
\end{eqnarray}

\twocolumngrid

\section{Results and discussion}
The above evaluations are the main analytic task we performed. It involves to identify several non-equivalent classes of dominant paths,  made of non-commuting sequences of blocks $X_{ij}$, which are dominant in the $N \to \infty$ limit, to evaluate their cardinality, to average over the random unit vectors in the $R^d$ space, and to average over the probability of a block to be non-zero. \\

Eqs.(3),(4) are displayed in a way to point out that the lowest moments are polynomials in the variable $Z/d$ whereas moments of higher order, starting with $\mu_8$ and $\nu_4$, have  additional terms involving just the space dimension $d$.\\
 
We proceed to compare these moments, with the moments of three limiting cases, as it is schematically indicated in  Fig.1.\\

Some relations are obvious but other are new and valuable.\\

First, in the $d\to 1$ limit our model reduces to the Erdos-Renyi graph. The moments of the spectral distributions of the Adjacency matrix  and Laplacian matrix were determined by recurrence relations at every order~\cite{bau}. Those moments are reproduced by setting  $d=1$ in
 Eqs.(3),(4)  and this is merely a consistency check of our evaluations.\\

A second limiting case is shown in Fig.1: the average connectivity $Z$ is allowed to increase as the order $N$ of the matrices increase: $Z/d \to \infty$ with $d$ fixed. In this limit, the number of non-zero blocks in each row of the matrices increases in the $N\to \infty$ limit, still keeping  $Z/N \to 0$. It is sometimes referred as the dilute matrix limit. Many investigations found that in this limit the spectral distribution of the matrix is the same as a symmetric matrix with independent entries. \\

\onecolumngrid

\begin{figure*}[h]
\begin{center}
\epsfig{file=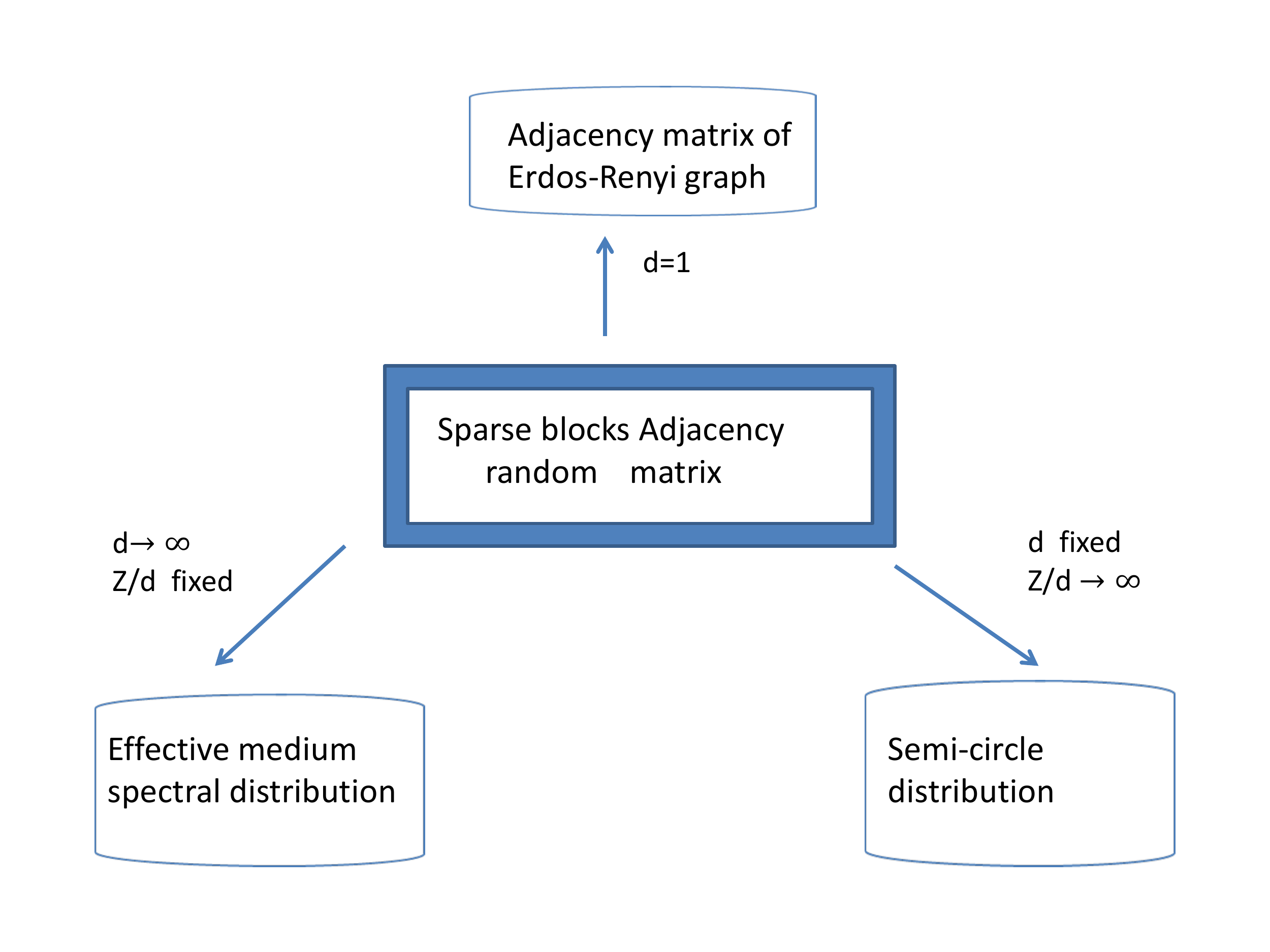, width=8.55cm  } \quad \epsfig{file=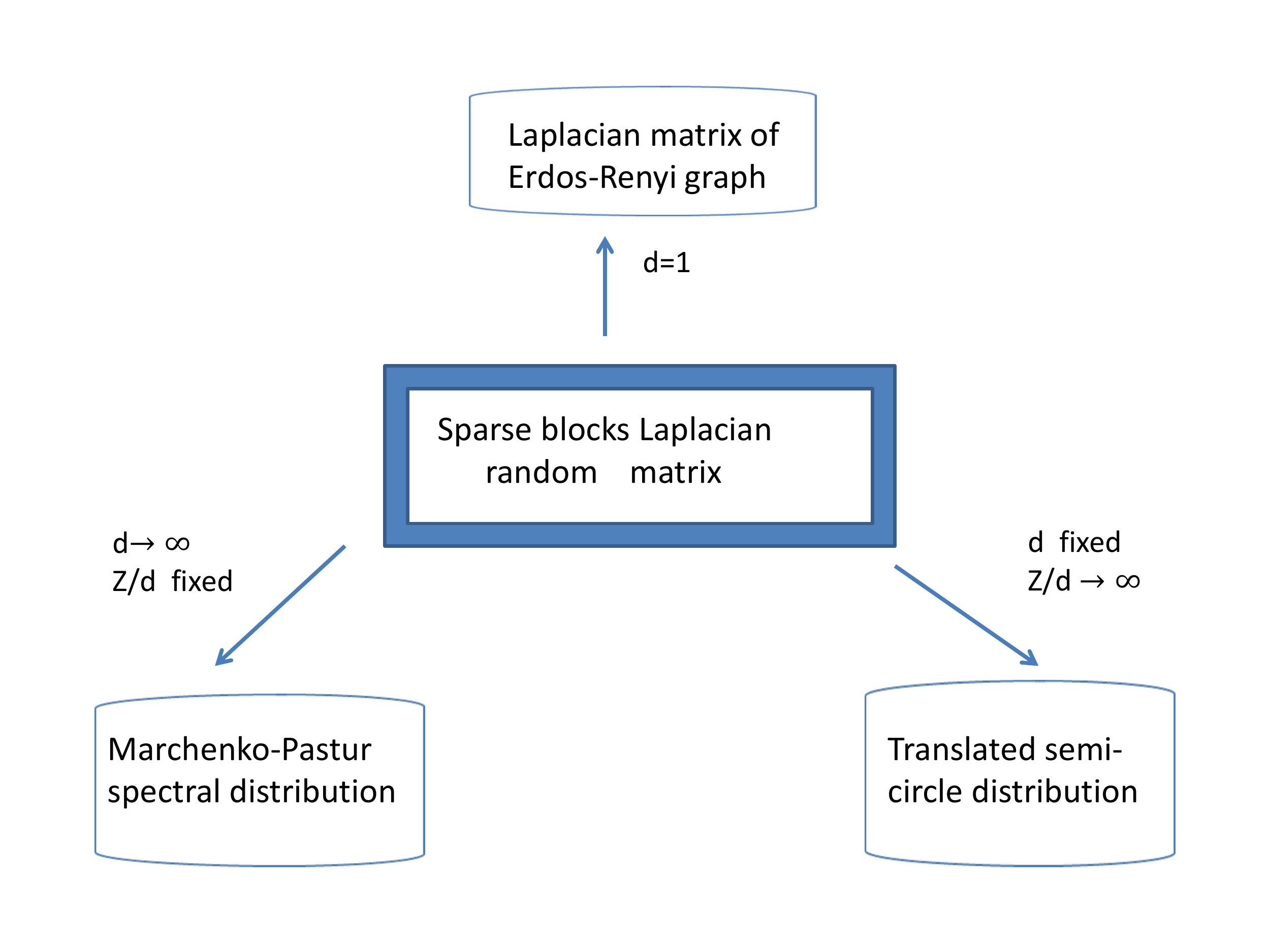, width=8.55cm  }
 \caption{The  left side shows the relation of the Adjacency block matrix with three simpler models in different limiting cases. The right side shows the parallel relations of the Laplacian block matrix.  \label{fig1}}
\end{center}
\end{figure*}

\twocolumngrid

Let us consider the Wigner semi-circle distribution and its well known moments (Catalan coefficients)
 \begin{eqnarray}
\rho(x)&=& \frac{\sqrt{4(Z/d)-x^2}}{2\pi(Z/d)} \quad , \quad -2\sqrt{Z/d}\leq x \leq 2\sqrt{Z/d} \nonumber\\
 \mu_{2k}&=& \frac{ (2k)!}{k!(k+1)!}\left(\frac{Z}{d}\right)^k \qquad
\label{d.7}
\end{eqnarray}
These moments reproduce the highest powers of the polynomials of Eq.(\ref{d.3}). Now let us consider the shifted semi-circle distribution and the first five moments
 \begin{eqnarray}
\rho(x)= \frac{1}{4\pi(Z/d)} \sqrt{ 8(Z/d)-(x-Z/d)^2}  \quad , \nonumber\\
 Z/d-2\sqrt{2(Z/d)}\leq x\leq Z/d+2\sqrt{2(Z/d)} \qquad , \nonumber\\
 \begin{array}{cccccccccccc}
& \nu_1 = &\frac{Z}{d} \\
& \nu_2 = &2\, \frac{Z}{d} &+ \left( \frac{Z}{d} \right)^2 \\
&\nu_3 = & &6\, \left( \frac{Z}{d} \right)^2 &+ \left( \frac{Z}{d} \right)^3  \\
&\nu_4 =  & &8\,\left( \frac{Z}{d} \right)^2  &+12\, \left( \frac{Z}{d}\right)^3 &+\left(\frac{Z}{d}\right)^4  \\
&\nu_5 = &  & &40\,\left( \frac{Z}{d}\right)^3   &+
20\,\left(\frac{Z}{d}\right)^4 &+ \left(\frac{Z}{d}\right)^5  
\end{array} \nonumber\\
\label{d.9}
\end{eqnarray}
These moments reproduce the leading and the first non-leading powers of the polynomials of Eq.(\ref{d.4}).\\

New and more relevant relations are related to the third limiting case: the limit $d\to \infty$ , for $Z/d$ fixed.\\
Semerjian and Cugliandolo \cite{semer} evaluated the effective medium (EM) approximation for the spectral distribution of the ensemble of  $N \times N$ real symmetric matrices where the diagonal elements vanish and the off-diagonal entry $J_{i,j}$ , $i<j$ is zero with probability $1-p/N$ and it is one with probability $p/N$:\\

\begin{widetext}
\begin{equation}
\rho(x)=\frac{\sqrt{3}}{2\pi}\left[ -\left(\frac{p-1}{3x}\right)^2-\frac{p+2}{6x}+\sqrt{\frac{(\lambda^2-x^2)(x^2+\alpha^2)}{27 x^4}}\right]^{1/3}\nonumber-\frac{\sqrt{3}}{2\pi}\left[ -\left(\frac{p-1}{3x}\right)^2-\frac{p+2}{6x}-\sqrt{\frac{(\lambda^2-x^2)(x^2+\alpha^2)}{27 x^4}}\right]^{1/3}\nonumber
\end{equation}
\end{widetext}

where $-\lambda\leq x\leq \lambda$, and $\lambda$, $\alpha$ are functions of $p$.\\

We evaluated  the moments of this spectral function from the Taylor expansion of the corresponding resolvent. One then obtains the moments in the table in Eq.(\ref{d.3}) where the terms $\frac{3}{d+2}$ are absent and $p=Z/d$. That is, the limit $d\to \infty$ with $Z/d$ fixed.\\

Finally, the same limit, $d\to \infty$, with $Z/d$ fixed, performed on the table in Eq.(\ref{d.4}) leads to

\begin{eqnarray}
\begin{array}{cccccccccccc}
& \nu_1 = \frac{Z}{d} \\
& \nu_2 = 2\, \frac{Z}{d} &+ \left( \frac{Z}{d} \right)^2 \\
&\nu_3 = 4\, \frac{Z}{d}& +6\, \left( \frac{Z}{d} \right)^2 &+ \left( \frac{Z}{d} \right)^3  \\
&\nu_4 = 8\, \frac{Z}{d}& +24\, \left( \frac{Z}{d} \right)^2  &+12 \left( \frac{Z}{d}\right)^3 &+\left(\frac{Z}{d}\right)^4  \\
&\nu_5 = 16\, \frac{Z}{d}& +80\, \left( \frac{Z}{d}\right)^2  &+80\,\left( \frac{Z}{d}\right)^3   &+
20\,\left(\frac{Z}{d}\right)^4 &+ \left(\frac{Z}{d}\right)^5 
\end{array}
\label{d.10}
\end{eqnarray}

The moments  $\int_a^b dx \, x^k\, \rho_{MP}(x)$ of the Marchenko-Pastur distribution 
 \begin{equation}
\rho_{MP}(x) = \frac{ \sqrt{(b-x)(x-a)}}{4 \pi\,x} \quad , \quad 0\leq a\leq x\leq b \nonumber\\
\end{equation}
with the following definition of parameters:
\begin{equation}
a=\left(\sqrt{p}-\sqrt{2}\right)^2 \quad, \quad b=\left(\sqrt{p}+\sqrt{2}\right)^2 
\nonumber
\end{equation}

where $p=Z/d$ reproduce the above Eq.(\ref{d.10}).\\

It is important to support the analytic indications of few moments with the full numerical evaluation of the spectral distributions. Large $Nd \times Nd $ block-Adjacency matrices and block-Laplacian matrices, with $N=1000 - 15000$ and $d=1,2,3,4,5,10,20$ were generated according the probability distribution of our model and the eigenvalues were numerically evaluated. The obtained spectral distributions are in Fig.2. They support the conjectured limits indicated in Fig.1 and the emerging unifying picture. 

Strikingly, while the difference between MP distribution and the numerical results for $d=3$ is of quantitative nature for the Laplacian, the difference between the EM approximation and the numerics for $d=3$ is of qualitative nature, especially around $\lambda=0$ where the numerical results for $d=3$ show a delta-like peak whereas EM predicts a saddle.\\

\section{Conclusions}
 In conclusion, the analytic evaluations of a few limiting moments and the numerical simulations support the conjecture of the relations schematically indicated in Fig.1 among different random matrix models. Since in the traditional models  of disordered systems through random matrices and  replica approach, the space dimension does not enter in the formulation of the model, the argument that the Effective Medium approximation (for the Adjacency matrix) and the Marchenko-Pastur  distribution (for the Laplacian matrix) are valid  for infinite space dimension is rather indirect and not well defined. The proposed relations and the systematic numerical results presented in this work substantiate these arguments by clarifying the role of space dimension for the various random matrix models, and suggest new ways to investigate disordered systems in finite space dimension.\\

\onecolumngrid

 \begin{figure}[h]
                \centering
                \includegraphics[trim=0cm 0.0cm 0cm
0.0cm,clip=true,width=500px,height=250px]{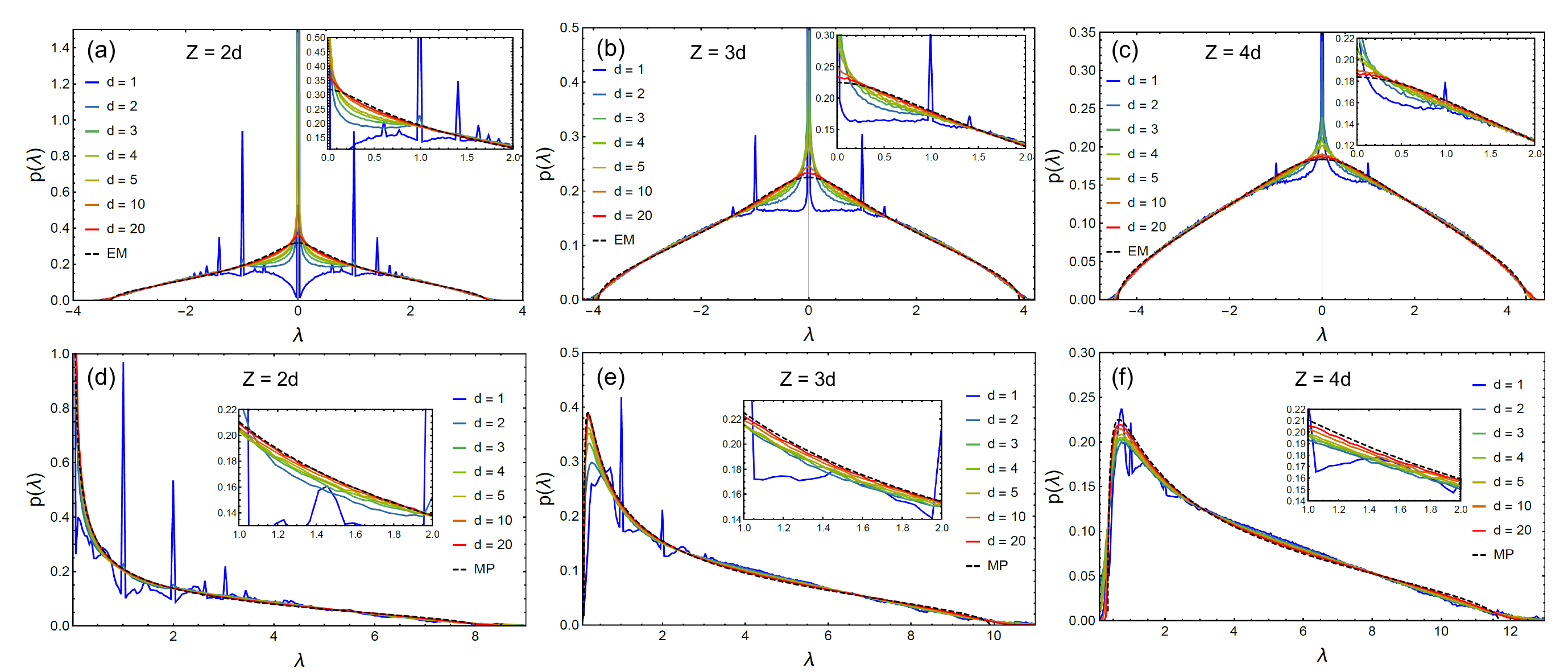}
                \caption{(a-c): Plots of the eigenvalue spectra of the
adjacency matrix obtained from our model systems for $d=1,2,3,4,5,10,20$ and
corresponding $N=15000, 7500, 5000, 4000, 3000, 2000, 1000$. They approach the
spectrum from Effective Medium Theory, which appears as the infinite
dimensional limit. (d-f): The spectra of the Laplace matrix for the same
systems. As one can see they approach the Marchenko-Pastur destribution for
infinite dimension.}
                \label{fig2}
\end{figure}

\twocolumngrid

 In regard to the theory of random matrices, the present model explores ensembles of blocks random matrices with two different probabilities: the probability of independent identically distributed blocks to occur and the probability of the entries in the blocks. This structure is new, very promising, and of great relevance for physics applications. \\
The conjectured  relations 
schematically indicated in Fig.1 indicate that this structure interpolates among all best studied spectral distributions.\\

 We are also confident that the limiting moments here evaluated will be useful in the search for suitable approximate analytic representations of the eigenvalue distributions of physical models in finite space dimensions.

\appendix
\section{Definition of the model}
It is useful to recall the well known correspondence between any  real symmetric matrix $M$ of order $N$ and the corresponding non-directed graphs $G$ with $N$ vertices. Between a generic pair of vertices $(i,j)$  of the graph there is a link, or edge, with the weight $M_{i,j}$. The edge is absent if the corresponding matrix entry is zero. Edges where the extrema of the edge is the same vertex correspond to the diagonal entries of the matrix.
 The matrix element of a power of the matrix, say $(M^k)_{i,j}$ may be evaluated as the sum of the contributions of weighted paths of $k$ steps from vertex $i$ and vertex $j$ on the graph.
$$(M^k)_{i,j} = \sum_{s_1=1,..,N,..,s_{k-1}=1,..N} M_{i,s_1}M_{s_1,s_2}\cdots M_{s_{k-1},j}$$

The sparse random block matrix we study in this work, is an ensemble of real symmetric matrices $M$ of dimension $Nd \times Nd$.\\
The generic matrix of the ensemble is a block matrix, with $N$ blocks in each row and column. Each block $X_{i,j}$ is a real symmetric matrix of order $d\times d$ 

\begin{eqnarray}
M=\left( \begin{array}{cccccccc}
X_{1,1} & X_{1,2} & X_{1,3}& \dots & X_{1,N}\\
X_{2,1}& X_{2,2} & X_{2,3} & \dots & X_{2,N}\\
\dots & \dots & \dots & \dots & \dots\\
X_{N,1} & X_{N,2}&X_{N,3}&\dots & X_{N,N} \end{array}\right)
  \qquad  \qquad
\label{a.1}
\end{eqnarray}

The blocks  $X_{i,j}$ are independent identically distributed random matrices.\\
 The  graph  corresponding to the matrix $M$ has $N$ vertices, the weight of the (non-directed) edge connecting the pair of vertices $(i,j)$ is  a $d\times d$ matrix $X_{i,j}=X_{j,i}=X_{i,j}^t$. It is still useful to evaluate elements of powers of the matrix in terms of the weighted paths connecting the vertices. Since the  weight of a path is a product of non-commuting blocks, the order of them is relevant.\\

The Adjacency matrix has a zero $d \times d$ block on the diagonal entries.
 \begin{eqnarray}
A=\left( \begin{array}{cccccccc}
 0 & X_{1,2} & X_{1,3}& \dots & X_{1,N}\\
X_{2,1}& 0 & X_{2,3} & \dots & X_{2,N}\\
\dots & \dots & \dots & \dots & \dots\\
X_{N,1} & X_{N,2}&X_{N,3}&\dots &  0 \end{array}\right)
  \qquad  \qquad
\label{a.2}
\end{eqnarray}

One easily evaluates traces of powers in terms of classes of  non-equivalent paths~\cite{cic}. Since the blocks are independent identically distributed random matrices, it is sufficient to record when a block has previously appeared in a path. Then $X_1$ stands for any of the $N(N-1)/2$ blocks $X_{i,j}$, $X_2$ stands for any block, different from $X_1$, etc. For instance
 \begin{eqnarray}
\frac{1}{N(N-1)}\mathrm{Tr}\,A^4&=& \mathrm{Tr}_d X_1^4+2\,(N-2)\,\mathrm{Tr}_d X_1^2X_2^2+\nonumber\\
&+&(N-2)(N-3)\,\mathrm{Tr}_d X_1X_2X_3X_4.
 \qquad  \qquad
\label{a.3}
\end{eqnarray}

The analogous evaluation for the Laplacian block matrix $L$ is more involved

\begin{widetext}
\begin{eqnarray}
L=\left( \begin{array}{cccccccc}
\sum_{j\neq 1}X_{1,j} & -X_{1,2} & -X_{1,3}& \dots & -X_{1,N}\\
-X_{2,1}& \sum_{j\neq 2}X_{2,j} & -X_{2,3} & \dots & -X_{2,N}\\
\dots & \dots & \dots & \dots & \dots\\
-X_{N,1} & -X_{N,2}& -X_{N,3}&\dots & \sum_{j\neq N}X_{N,j} \end{array}\right)
  \qquad  \qquad
\label{a.4}
\end{eqnarray}
\begin{eqnarray}
\frac{1}{N(N-1)}\mathrm{Tr}\,L^4&=& 8\, \mathrm{Tr}_d X_1^4+16\,(N-2)\,\mathrm{Tr}_d X_1^3X_2\nonumber\\
&+& 8\,(N-2) \mathrm{Tr}_d X_1^2X_2^2 \nonumber\\
&+&\,(N-2)\,\mathrm{Tr}_d X_1X_2X_1X_2\nonumber\\
&+&8\,(N-2)(N-4)\,\mathrm{Tr}_d X_1^2X_2X_3 \nonumber\\
&+&\,2\,(N-2)(2N-5)\,\mathrm{Tr}_d X_1X_2X_1X_3 \nonumber\\
&+& (N-2)(N-3)(N-7)\,\mathrm{Tr}_d X_1X_2X_3X_4
 \qquad  \qquad
\label{a.5}
\end{eqnarray}
\end{widetext}

Each block $X$ is the null matrix $d\times d$, with probability $1-(Z/N)$ or it is a rank-one random matrix $X=\hat{n}\hat{n}^t$, with probability $Z/N$, where $\hat{n}$ is a random vector of length one, chosen with uniform probalibilty in $R^d$.\\

 Then, for instance,
$\mathrm{Tr}_d X_1X_2X_3X_4=0$ with probability $1-(Z/N)^4$ or $(\hat{n}_1\hat{n}_2)(\hat{n}_2\hat{n}_3)(\hat{n}_3\hat{n}_4)(\hat{n}_4\hat{n}_1)$ with probability $(Z/N)^4$.
And  $\mathrm{Tr}_d X_1X_2X_1X_3=0$ with probability $	1-(Z/N)^3$ or $(\hat{n}_1\hat{n}_2)^2(\hat{n}_3\hat{n}_1)^2
$ with probability $(Z/N)^3$. The expected number of non-zero $d\times d$ blocks  in each row or column of the Adjacency matrix is $\frac{N-1}{N}Z$,  then $Z$ is the average connectivity of the large graph (or the average degree of the vertices).\\

 Finally the average over the uniform probability of the direction of all the random vectors $\hat{n}_j$ involves integrals for each of them over the unit sphere in $R^d$. Let us denote $<...>_d$ such integrals. For instance
\begin{eqnarray}
&&< (\hat{n}_1\hat{n}_2)^2>_d=\frac{1}{d} \quad , \quad < (\hat{n}_1\hat{n}_2)^4>_d=\frac{3}{d(d+2)} \quad , \nonumber\\
&&< (\hat{n}_1\hat{n}_2)^6>_d=\frac{15}{d(d+2)(d+4)} \quad , \nonumber\\
&& <(\hat{n}_1\hat{n}_2)(\hat{n}_2\hat{n}_3)(\hat{n}_3\hat{n}_1)>_d=\frac{1}{d^2}\quad , \nonumber\\
&& <(\hat{n}_1\hat{n}_2)(\hat{n}_2\hat{n}_3)(\hat{n}_3\hat{n}_4)(\hat{n}_4\hat{n}_1)>_d=\frac{1}{d^3}
 \qquad  \qquad
\label{a.6}
\end{eqnarray}

By this method we find from Eqs.(\ref{a.3}) and (\ref{a.5})
\begin{eqnarray}
\lim_{N \to \infty}\frac{<\mathrm{Tr}A^4>}{Nd}&=&\frac{Z}{d}+2\left(\frac{Z}{d}\right)^2 \quad , \nonumber\\
\lim_{N \to \infty}\frac{<\mathrm{Tr}L^4>}{Nd}&=&8\frac{Z}{d}+\left(\frac{Z}{d}\right)^2 \left(24+\frac{3}{d+2}\right) \nonumber\\
&+& 12\,\left(\frac{Z}{d}\right)^3+ \left(\frac{Z}{d}\right)^4.
  \nonumber
	\end{eqnarray}


\section{The moments of the limiting models}
\subsection{The simple random graph} 
For a simple (that is: no multiple edges, no edge with just one vertex) random graph, where the probability of any edge is $Z/N$, the moments of the spectral distribution of the Adjacency matrix and the Laplacian matrix were evaluated in the $N \to \infty$ limit, and fixed average connectivity $Z$ at every order \cite{bau}. We report here the first few moments, from Table 1 and 2 of Bauer and Golinelli~\cite{bau}.
For the Adjacency matrix we have:

\begin{equation}
\mu_k=\lim_{N \to \infty}\frac{1}{N} <{\rm Tr}\,A^k>  \quad , \quad \mu_0=1\quad , \quad   \mu_{2k+1}=0  \nonumber\\
\end{equation}
which produces:

\begin{eqnarray}
\begin{array}{ccccccc}
&\mu_2= Z   \\
&\mu_4=Z &+2Z^2 \\
&\mu_6=Z &+6Z^2 &+5Z^3 \\
&\mu_8=Z &+14 Z^2 &+28 Z^3 &+14 Z^4  \\
&\mu_{10}=Z&+30\,Z^2&+110\, Z^3&+120\, Z^4&+42\, Z^5 \\
&\mu_{12}=Z &+62\,Z^2 &+375\,Z^3 &+682\,Z^4 &+495\,Z^5 &+132\,Z^6
\end{array}
\nonumber
\end{eqnarray}

while for the Laplacian matrix we have:

\begin{equation}
\nu_k =\lim_{N \to \infty}\frac{1}{N} <{\rm Tr}\,L^k> \quad , \quad \nu_0=1 \nonumber\\
\end{equation}
which produces:

\begin{eqnarray}
\begin{array}{cccccccccc}
&\nu_1= Z \\
&\nu_2= 2\,Z  &+Z^2\\
&\nu_3=4\, Z &+ 6\, Z^2& +Z^3 \\
&\nu_4= 8\,Z &+25\, Z^2&+12\, Z^3& +Z^4 \\
&\nu_5=  16\,Z&+90\,Z^2&+85\,Z^3&+20\, Z^4&+Z^5 \\
&\nu_6=  32\,Z&+301\, Z^2&+476 \,Z^3&+ 215\, Z^4&+30\, Z^5&+Z^6.
\end{array}
\nonumber
	\end{eqnarray}
	\vskip 1 cm

\subsection{Effective medium approximation} 
 In the same model, the spectral distribution of the Adjacency matrix in the Effective Medium (EM) approximation, is

	\begin{eqnarray}
	\rho^{\mathrm{EM}}(x) &=&-\frac{1}{\pi}\mathrm{Im}\,g(x+i\epsilon) \quad , \quad g(z)= \int \frac{\rho^{\mathrm{EM}}(x) }{z-x}\,dx\nonumber\\
\rho^{\mathrm{EM}}(x) &=& \frac{\sqrt{3}}{2\pi}\left[ \left(\frac{p-1}{3 x}\right)^2+\frac{p+2}{6 x}+\sqrt{ \frac{(\lambda^2-x^2)(x^2-\alpha^2)}{27 x^4} } \right]^{1/3}\nonumber\\
&-&\frac{\sqrt{3}}{2\pi}\left[ \left(\frac{p-1}{3 x}\right)^2+\frac{p+2}{6 x}-\sqrt{ \frac{(\lambda^2-x^2)(x^2-\alpha^2)}{27 x^4} } \right]^{1/3}\nonumber
		\end{eqnarray}
	where $-\lambda\leq x\leq \lambda$ 
\begin{eqnarray}
\lambda &=& \sqrt{\frac{-p^2+20 p+8+\sqrt{p(p+8)^3}}{8}} \nonumber\\
\alpha^2 &=& \frac{p^2-20 p-8+\sqrt{p(p+8)^3}}{8}
\nonumber
	\end{eqnarray}
It is difficult to evaluate  the moments $\mu_{2k}=\int_{-\lambda}^\lambda x^{2k} \,\rho^{\mathrm{EM}}(x)\,dx$ by analytic integration, but the first few moments are easily obtained from the series solution of the cubic
	\begin{eqnarray}
&& [g(z)]^3+\frac{p-1}{z} \,[g(z)]^2-g(z)+\frac{1}{p}=0 \quad , \quad \nonumber\\
&& g(z)=\sum_{k=0}^\infty \frac{\mu^{2k} }{z^{2k+1}}=\frac{1}{z}+\frac{p}{z^3}+\frac{p+2p^2}{z^5}+\frac{p+6p^2+5p^3}{z^7}+    \nonumber\\
&&\quad +\frac{p+11p^2+28p^3+14p^4}{z^9}+ \frac{p+20p^2+90p^3+120p^4+42p^5}{z^{11}}+  \nonumber\\
&&\quad +\frac{p+30p^2+220p^3+550p^4+495p^5+132p^6}{z^{13}}+\nonumber\\
&&\quad +\frac{p+42p^2+455p^3+1820p^4+3003p^5+2002p^6+429p^7}{z^{15}}\\
&&\quad +O(z^{-17}).
\nonumber
	\end{eqnarray}

\subsection{Marchenko-Pastur distribution}
The  Marchenko-Pastur distribution reads as
 \begin{eqnarray}
\rho_{MP}(x) &=& \frac{  \sqrt{(b-x)(x-a)}}{4 \pi\,x} \quad , \quad 0\leq a\leq x\leq b \nonumber\\
a&=&\left(\sqrt{p}-\sqrt{2}\right)^2 \quad , \quad b=\left(\sqrt{p}+\sqrt{2}\right)^2.
\nonumber
\end{eqnarray}
The moments  $ \nu_k=\int_a^b dx \, x^k\, \rho_{MP}(x)$  are well known, and are given by
 \begin{eqnarray}
\nu_k&=&\int_a^b dx \, x^k\, \frac{  \sqrt{(b-x)(x-a)}}{4 \pi\,x} = \nonumber\\
&=&\frac{ (2p)^{(k+1)/2} 2^{k-1}}{\pi}\int_{-1}^1\left(t+\frac{p+2}{\sqrt{8p}}\right)^{k-1}\sqrt{1-t^2}\,dt = \nonumber\\
&=&p(p+2)^{k-1}\,_2F_1 \left(\frac{1-k}{2}, 1-\frac{k}{2};2;\frac{8p}{(p+2)^2}\right).
\nonumber
	\end{eqnarray}

\end{document}